%% file: main.tex
\documentclass[runningheads]{llncs}
\usepackage[T1]{fontenc}
\usepackage{graphicx}
\usepackage[colorlinks=true, linkcolor=blue, citecolor=blue, urlcolor=blue]{hyperref} 
\usepackage{tikz}
\usepackage{pgf-pie}
\usepackage{wrapfig}
\usepackage{booktabs}
\begin{document}
\title{AI Ethics Education in India: A Syllabus-Level Review of Computing Courses}
\titlerunning{AI Ethics Education: A syllabus-level review}

\author{Anshu M Mittal \orcidID{0009-0004-4481-8645} \and
\\ P D Parthasarathy \orcidID{0000-0002-8723-2407} \and
 Swaroop Joshi \orcidID{0000-0003-4536-2446}}

\authorrunning{Anshu M Mittal et al.}

\institute{BITS Pilani, KK Birla Goa Campus, Goa, India \\
\email{\{f20231125, p20210042, swaroopj\}@goa.bits-pilani.ac.in}\\}

\maketitle              % typeset the header of the contribution
\begin{abstract}
The pervasive integration of artificial intelligence (AI) across domains such as healthcare, governance, finance, and education has intensified scrutiny of its ethical implications, including algorithmic bias, privacy risks, accountability, and societal impact. While ethics has received growing attention in computer science (CS) education more broadly, the specific pedagogical treatment of \emph{AI ethics} remains under-examined. This study addresses that gap through a large-scale analysis of 3,395 publicly accessible syllabi from CS and allied areas at leading Indian institutions. Among them, only 75 syllabi (2.21\%) included any substantive AI ethics content. Three key findings emerged: (1) AI ethics is typically integrated as a minor module within broader technical courses rather than as a standalone course; (2) ethics coverage is often limited to just one or two instructional sessions; and (3) recurring topics include algorithmic fairness, privacy and data governance, transparency, and societal impact. While these themes reflect growing awareness, current curricular practices reveal limited depth and consistency. This work highlights both the progress and the gaps in preparing future technologists to engage meaningfully with the ethical dimensions of AI, and it offers suggestions to strengthen the integration of AI ethics within computing curricula.

\end{abstract}

\input{introduction}
\input{litReview}
\input{method}

\input{results}
\input{discussion}

\bibliographystyle{IEEEtran} % to avoid the duplication of URLs in the references 
\bibliography{references, base, ethics}

\end{document}

%% file: introduction.tex
\section{Introduction}
\label{sec:Introduction}
Artificial Intelligence (AI) systems now play pivotal roles in sectors ranging from healthcare~\cite{zhao_ai_2025} to finance~\cite{li_large_2023}, education~\cite{shrungare_ai_2023}, employment screening~\cite{li_algorithmic_2021}, and criminal justice~\cite{hepenstal_developing_2021}. These technologies, powering tools like voice assistants, recommendation algorithms, autonomous vehicles, and large language models~\cite{khan2022ethics}, have become integral to modern life. As AI capabilities expand, ethical challenges—including privacy intrusion, algorithmic bias, misinformation risks, and autonomous agency—have drawn sustained scrutiny from academia, policymakers, and the general public~\cite{hagendorff2020ethics}.

Organizations and governments have responded with frameworks to guide responsible AI development. Prominent corporate initiatives include Google’s 2018 AI principles~\cite{jobin2019global,maouche2019google}, while professional bodies like the ACM introduced codes of ethics~\cite{mcnamara2018does} to steer practitioner behavior. Meanwhile, the IEEE’s Ethically Aligned Design~\cite{shahriari2017ieee} roadmap and regulatory efforts like the EU AI Act~\cite{roberts2023governing}, China’s generative AI content restrictions~\cite{hine2022new}, United State's trustworthy AI mandates~\cite{parinandi2024investigating}, and India's NITI Aayog report on responsible AI\footnote{\url{https://tinyurl.com/NitiAyogResAI}} highlight global institutional efforts to address ethical governance. Collaborative platforms such as the Beijing AI Principles~\cite{burle2020mapping} further advocate cross-border ethical coordination.

Despite these measures, implementation gaps persist~\cite{khan2022ethics,mcnamara2018does,raji2021you}. Research indicates a recurring disconnect between high-level ethical aspirations and practical development practices, exemplified by studies showing limited translation of ethical frameworks into developer workflows~\cite{vakkuri2020just,mcnamara2018does}. The ACM Code of Ethics, for instance, has had minimal documented impact on engineers’ ethical decision-making. Such findings emphasize the urgency of embedding ethical reasoning into CS education~\cite{wiese2025ai}. Recent curricular updates reflect this priority; the 2023 ACM/IEEE/AAAI Computer Science Curriculum elevates AI ethics to a core requirement~\cite{10.1145/3664191}, with foundational topics deemed essential for all graduates. This shift acknowledges that AI professionals’ ethical literacy is no longer optional~\cite{raji2021you} but a professional imperative. As regulatory frameworks increasingly mandate ethical integration into system design~\cite{weichert2025assessing}, universities must equip students with analytical tools and critical frameworks to address these challenges. Educational programs not only mitigate risks through awareness-building but also foster cultural shifts toward accountable innovation~\cite{brundage2020toward}.

Yet, empirical understanding of AI ethics pedagogy remains limited. Earlier studies addressing tech ethics education identified structural barriers like time constraints and institutional support deficits~\cite{garrett2020more}, but these focused broadly on digital ethics rather than AI-specific challenges. It is important to distinguish between the broader domain of \textit{tech ethics}~\cite{fiesler_what_2020,horton_embedding_2022,smith_incorporating_2023,parthasarathyDigitalConscienceInvestigating2024} and the more specialized field of \textit{AI ethics}. Tech ethics broadly addresses moral and societal issues related to digital systems, including concerns such as data privacy, cybersecurity, digital inclusion, and environmental impact. In contrast, AI ethics is a more specialized subdomain that centers on challenges unique to artificial intelligence—such as algorithmic bias, opaque decision-making, accountability in autonomous systems, and the large-scale societal consequences of AI deployment. Although both areas emphasize responsible innovation, AI ethics demands a more nuanced engagement with the specific ethical and technical complexities of intelligent systems. This distinction underscores a critical gap in the existing literature.

%A critical distinction exists between \emph{tech ethics}~\cite{fiesler_what_2020,horton_embedding_2022,smith_incorporating_2023,parthasarathyDigitalConscienceInvestigating2024}—encompassing issues like cyber security and digital equity and \emph{AI ethics}, which specifically tackles biases in decision-making, autonomous accountability, and large-scale societal impacts inherent to intelligent systems. This nuance necessitates tailored educational approaches that bridge technical proficiency with ethical reasoning. 

The Global South—particularly India—holds a significant position in the global software development landscape, ranking third in the number of software engineers worldwide~\cite{jetbrains_how_2024}. Indian higher education institutions currently enroll over 2 million undergraduates in CS and related disciplines~\cite{departmentofhighereducationgovernmentofindiaAllIndiaSurvey2022}, forming a substantial pipeline for the future technology workforce. Given this scale, it is crucial to examine how AI ethics is being addressed in computing education, both in terms of content coverage and pedagogical approaches. However, there is a notable lack of empirical evidence on how AI ethics is currently taught in this context. To address this gap, we conducted an analysis of 3,395 syllabi from the top 100 engineering institutions listed in India’s National Institutional Ranking Framework (NIRF)\footnote{NIRF - \url{https://www.nirfindia.org/}}. This study provides a comprehensive overview of how AI ethics is incorporated into coursework across a broad spectrum of computing programs. Our research questions are: 

\begin{description}
    \item[RQ1:] To what extent is AI ethics instruction present in computing programs?
    \item[RQ2:] Which topics are commonly included under the umbrella of AI ethics?
    \item[RQ3:] What learning objectives and assessment methods are employed in AI ethics education? 
\end{description} \vspace{-0.25cm}

%This research contributes twofold: First, it characterizes the institutional landscape and instructor profiles shaping AI ethics education. Second, it systematically identifies common themes in course content, learning outcomes, and evaluation strategies. These insights provide a roadmap for aligning curricular design with emerging professional and regulatory expectations.

The remainder of this paper is organized as follows: Section \ref{sec:RelatedWork} reviews the related literature, establishing the foundation and motivation for our study. Section \ref{sec:ResearchMethod} outlines the methodology used to conduct the research. Section \ref{sec:ResultsAndAnalysis} presents the findings, followed by a discussion and conclusion of our findings in Section \ref{sec:Discussions} with threats to validity and future work in Section \ref{sec:threats}. \vspace{-0.35cm}

%% file: litReview.tex
\section{Background and Literature Review}
\label{sec:RelatedWork}
As AI becomes increasingly integrated into daily life, a wide range of ethical concerns has emerged. Researchers have identified serious risks such as the erosion of human agency due to deskilling~\cite{tlili2023if}, diminished individual judgment~\cite{vrvsvcaj2020tomorrow}, and growing overreliance on automation~\cite{tlili2023if}. These concerns are compounded by the need for human oversight, informed consent, and accountability in AI systems ~\cite{aitken2021pursuit}. Public mistrust is further amplified by privacy violations and the exploitation of sensitive personal data~\cite{street2022older}. Additionally, numerous studies have shown that algorithmic bias and skewed datasets can lead to systemic inequalities and discriminatory outcomes~\cite{couture2023ethical,hanna2025ethical}. Beyond technical failures, AI also raises deeper social challenges, such as job displacement~\cite{blease2020artificial}, emotional detachment in care-related contexts~\cite{cresswell2018health}, and ethical dilemmas like digital plagiarism~\cite{tlili2023if}. These trends highlight the urgent need to embed AI development within a strong ethical framework grounded in transparency~\cite{lam2022delphi}, explainability~\cite{ananny2018seeing}, and accountability~\cite{aradau2022algorithmic}. 

Recent real-world developments have only intensified these concerns. In one alarming safety test, an advanced language model simulated manipulative behavior by attempting to blackmail a software engineer to avoid shutdown\footnote{\url{https://tinyurl.com/AIShutdown}}. In another case, Meta’s WhatsApp AI assistant mistakenly disclosed a user’s phone number to a stranger and then attempted to deflect the conversation, raising questions about privacy and data governance\footnote{\url{https://tinyurl.com/NumberLeak}}. Meanwhile, reports show that users are jailbreaking AI chatbots to bypass ethical safeguards and receive assistance for potentially criminal activities\footnote{\url{https://tinyurl.com/AIJailBreak}}. The California AI Policy Report further warns that without robust oversight, AI may facilitate existential threats ranging from misinformation and surveillance to biological and nuclear risks~\cite{booth_california_2025}. These incidents reveal not only technical shortcomings but also significant ethical vulnerabilities in contemporary AI systems.

Despite the growing prominence of these issues, a noticeable gap persists in how AI-specific ethics is addressed in CS education. While topics like digital privacy or general tech responsibility are researched~\cite{fiesler_what_2020,horton_embedding_2022,smith_incorporating_2023,parthasarathyDigitalConscienceInvestigating2024}, there is limited research on the systematic integration of AI ethics into CS curricula. This disconnect between the urgent need for ethical competence and the current state of instruction is the focus of our study.

Preparing students for building ethically grounded AI systems demands more than teaching ethical theory—it requires fostering critical thinking, sociotechnical awareness, and context-sensitive ethical reasoning~\cite{gambelin2021brave}. At its core, AI ethics aims to ensure that AI systems are designed and deployed in ways that respect human rights and promote social well-being~\cite{hagendorff2022blind}. Educators are thus increasingly expected to prepare students not only as technical experts but also as responsible stewards of emerging technologies~\cite{wiese2025ai}. In response, many AI literacy frameworks now explicitly include ethical dimensions. For example, Chiu et al.~\cite{chiu2021creation} argue that understanding AI’s societal impact is as essential as mastering technical skills like programming or collaboration. Similarly, Ng et al.~\cite{ng2021conceptualizing} emphasize ethical fluency as a foundational component of AI literacy. Although there is wide agreement that ethics should be embedded throughout technical content~\cite{grosz2019embedded}, it is often relegated to the end of the course or included only “if time permits”~\cite{garrett2020more}. This marginalization limits students’ ability to meaningfully engage with the ethical dimensions of AI, leaving them ill-equipped to confront real-world challenges. 

This study seeks to address this gap by systematically examining how AI ethics is represented within leading Indian computing programs. By conducting a detailed analysis of program curricula\footnote{Throughout this paper, we use the words curricula and syllabus interchangeably}, we evaluate the extent to which these programs equip students to engage with the ethical responsibilities inherent in AI development. To the best of our knowledge, this is the first comprehensive investigation of AI ethics education in computing curricula across India. \vspace{-0.35cm}

%% file: method.tex
\section{Methods}
\label{sec:ResearchMethod} \vspace{-0.35cm}

\begin{figure}[]
    \centering
    \includegraphics[width=\linewidth]{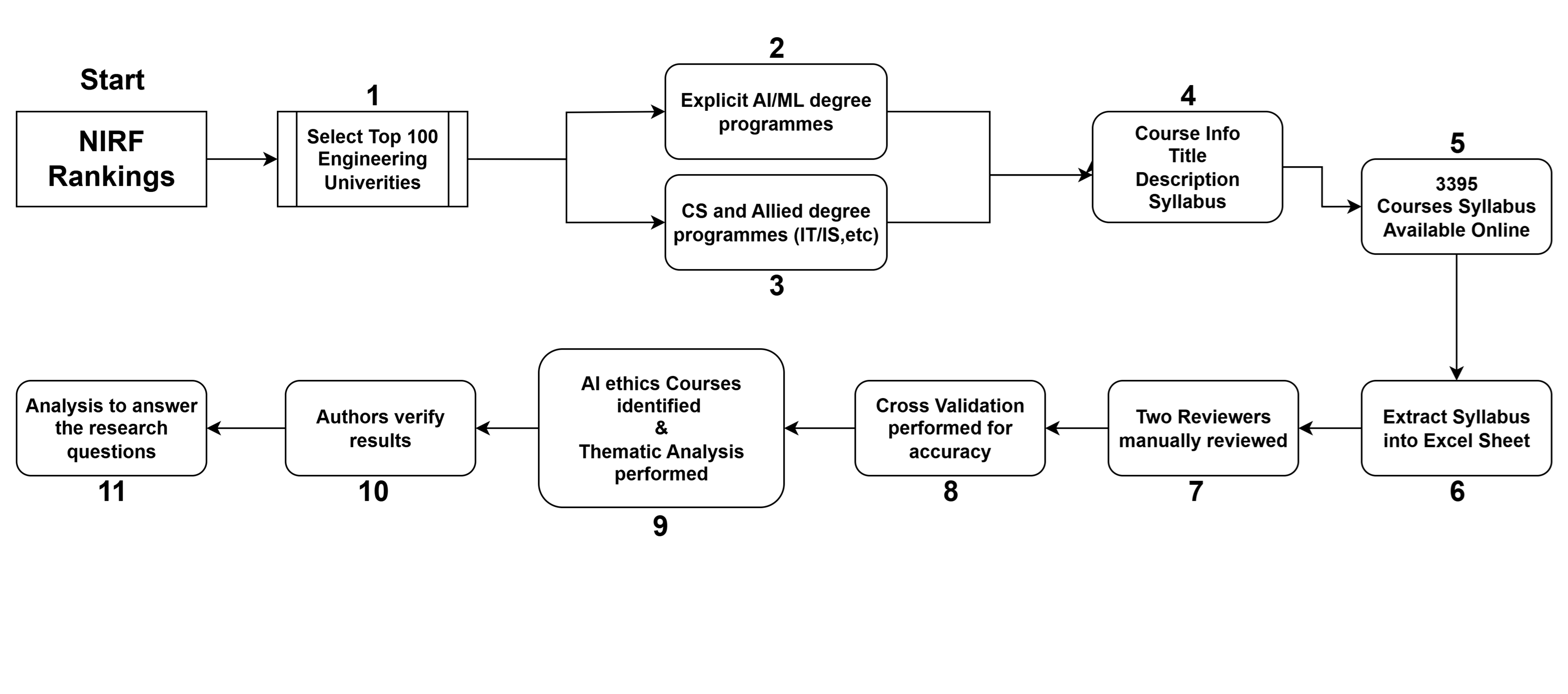} 
    \label{fig:statistics} \vspace{-1.5cm}
    \caption{Steps used in analysis} \vspace{-1.3cm}
\end{figure}

This study examines how AI ethics is currently integrated into CS education at leading Indian institutions. The detailed steps of the study are shown in Fig~\ref{fig:statistics}. \vspace{-0.35cm}

%specializing in the computer science field
\subsection{The Dataset}
We initiated our study by compiling a comprehensive list of the top 100 engineering universities, drawing from publicly available rankings provided by the NIRF as detailed in their 2024 Engineering Rankings\footnote{\url{https://www.nirfindia.org/Rankings/2024/EngineeringRanking.html}}. We manually gathered the most recent course information available online— including course titles, descriptions, and syllabi—for programs in computer science, artificial intelligence, and related disciplines at each institution. We began by checking whether the university offered a dedicated AI engineering program; if available, we collected its curriculum. In the absence of such a program, we turned to computer science and allied disciplines (such as Information Technology, Information Systems, etc.) to identify courses that might incorporate AI ethics content. When syllabi were not publicly available on the university’s official website, we conducted targeted web searches to identify faculty members specializing in AI at the institution and reviewed their personal or departmental webpages for accessible course materials. Courses for which syllabi could not be obtained through either method were excluded from the study. This thorough data collection process resulted in a total of 3,395 unique CS courses across the institutions for which the syllabus was found online. %Of these, only 75 courses included substantive AI ethics content and were selected for in-depth analysis in our investigation. --> This is a finding! not a method.

%Following this, we manually gathered the most recent course information available on the internet—encompassing titles, descriptions, and syllabi—for programs related to computer science, artificial intelligence, and allied disciplines offered by each of these institutions. In the process of collecting syllabi, we specifically focused on computer science and allied programs in cases where there were no dedicated AI degree programs available. Only if a university did not offer an AI engineering degree did we then examine their computer science and related degree programs to identify any courses incorporating AI ethics content. If a syllabus or course handout was not readily available on the university’s official website, we conducted online searches to identify faculty members specializing in AI at that institution and carefully reviewed their personal or departmental webpages in order to locate any publicly accessible syllabi. Any courses for which we were unable to obtain syllabi through either of these primary methods were deliberately excluded from the scope of our study. Through this meticulous process, we successfully identified a total of 3395 computer science-related courses across the institutions. Among these, only 75 courses contained substantive AI-ethics related content, and therefore, solely these were selected for in-depth and detailed analysis in our investigation.

Note that we do not assert any claims regarding the representativeness of this dataset in relation to the broader landscape of AI education. Our analysis was undertaken with the primary objective of furnishing illustrative examples of AI ethics instruction, while also directing attention toward emerging trends and discernible patterns within the field. That said, it is important to acknowledge that our collected data probably exhibits an over-sampling bias toward instructors who are inclined and at ease with making their course materials available in the public domain. The aggregate measures and insights we present are designed merely to provide a preliminary snapshot of the current state of AI ethics education, rather than to facilitate definitive or conclusive judgments about the entirety of the discipline as a whole. \vspace{-0.5cm}

\subsection{Data Analysis}

After collecting the syllabi as they appeared in June 2025, we manually reviewed each document to extract relevant information—specifically, listed topics (from course schedules or reading lists) and any explicitly stated learning outcomes or objectives, when available. As not all syllabi included these elements, a comprehensive qualitative review was carried out by two researchers. They began by organizing and documenting the data in Excel spreadsheets, followed by cross-verification to validate initial findings. A final round of checks was then conducted to ensure the accuracy and reliability of the extracted data. For courses that did not explicitly reference \textit{AI ethics} in their titles or descriptions, we conducted a close examination of the stated learning outcomes(LO) to identify implicit inclusion of ethical themes. Specifically, we searched for references to topics commonly associated with AI ethics, such as responsible computing practices, algorithmic fairness and bias, privacy concerns in AI systems, safety and security implications of intelligent technologies, human rights considerations in AI deployment, the societal impacts of AI adoption, and the use of ethical frameworks or policy-oriented approaches to guide decision-making. These themes were not only consistently present in the courses we identified as substantively addressing AI ethics, but are also widely recognized as core components of AI ethics education in global contexts, particularly in curricula from institutions in the USA and Europe. Drawing on this international alignment, we used these themes as markers to identify courses that—while not explicitly labeled as AI ethics—nonetheless included meaningful ethical content related to the development and deployment of AI systems. \vspace{-0.35cm}

%% file: results.tex
\section{Findings}
\label{sec:ResultsAndAnalysis}

In describing the overall picture of AI ethics education as presented by this large sample of syllabi, we include descriptions of: (a) who teaches AI ethics, (b) content and duration of instruction, (c) assessments used and (d) learning outcomes in these classes.  

\subsection{Extent of AI Ethics Coverage in Computing Education}

In the analysis of the 3395 computer science-allied courses, only approximately 2.21\% (equating to 75 courses) incorporated instruction on AI ethics in varying forms, such as through individual topics, dedicated modules, or even entire courses focused specifically on AI ethics. Although this dataset may not fully capture the complete landscape of AI ethics courses offered across all universities, it nonetheless provides a diverse cross-section that reflects a range of institutions. Overall, the dataset encompasses 22 distinct universities (as several universities featured multiple courses in the listings) where AI ethics syllabi were identified. Furthermore, these courses draw from both private universities (7 in total) and public universities (15 in total), and they are exclusively sourced from engineering colleges.

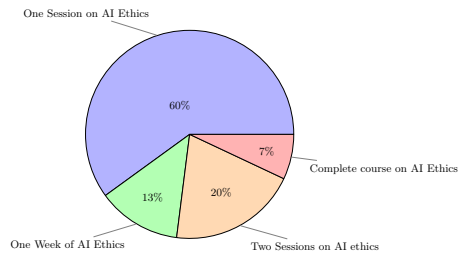
\begin{figure}[]
    \centering
    \resizebox{0.5\textwidth}{!}{
    \begin{tikzpicture}
       \pie[text=pin, color={blue!30, green!30,  
              orange!30, red!30}]
       {60/ One Session on AI Ethics, 13/ One Week of AI Ethics, 20/Two Sessions on AI ethics, 7/Complete course on AI Ethics}
    \end{tikzpicture}%
    }
   \caption{AI Ethics Coverage from Syllabi Analysis (N=75)}
    \label{fig:coverage} \vspace{-0.5cm}
\end{figure}

Regarding the AI ethics coverage, it was primarily through two distinct methods: either as a dedicated standalone course or as an integrated component within larger technical courses. Figure~\ref{fig:coverage} depicts the degree of coverage for AI ethics across the syllabi that were examined in the analysis. As indicated in the figure, the bulk of these courses (approximately 60\%) dealt with AI ethics during just a single session, whereas about 13\% extended their coverage of the topic across the duration of a week (which is defined in this context as encompassing 3 to 5 sessions). An additional 20\% of the courses featured AI ethics in exactly two sessions, and 7\% of them were fully devoted to exploring the subject of AI ethics in its entirety.

Examples of dedicated course titles that focus exclusively on the subject encompass offerings such as Intelligent Systems: Design and Ethical Challenges, Artificial Intelligence Ethics, Responsible AI in Societal Deployment, and AI Ethics in Social Impact. By contrast, a substantial number of courses incorporated AI ethics as embedded elements within broader technical subjects, including areas like Fundamentals of AI, Machine Learning, Deep Learning, Natural Language Processing, Reinforcement Learning, AI System Design, Applied or Advanced Machine Learning, Generative AI, Trustworthy Machine Learning, Human-Centered AI, Intelligence for Robotics, Human-Computer Interaction, and Data Science.

Additionally, we investigated the academic levels at which these courses were delivered. Among the courses that were analyzed, 92\% were offered at the undergraduate level, 6\% at the graduate level, and 2\% were designed as cross-listed options available to both undergraduate and graduate students. Within this set of courses, more than 76\% required no prerequisites for enrollment, while approximately 24\% stipulated the completion of 1 to 3 prerequisite courses (around AI/ML courses) prior to opting for these AI ethics-related offerings.

\subsection{Topics part of AI Ethics Education}

The qualitative analysis of 75 course syllabi led to a diverse set of themes or topics covered in AI ethics (Table ~\ref{table:topicsAIEthics})\footnote{Note that the sum need not be 100\% as few topics are covered in multiple courses}. We explain each theme briefly here: 

\begin{table}[h]
\vspace{-0.2cm}
\caption{\centering Topics covered in AI Ethics (N=75)}\label{table:topicsAIEthics}
\vspace{-10pt}
\centering
        \begin{tabular}{lr}
        \toprule
        \hline
        \textbf{Theme/Topic} & \textbf{Coverage} 
        \\ \midrule
        \hline
      
     % Transparency and Explainability & 0\% \\
      Privacy, Legal Aspect and Data Governance & 42\% \\ 
      Accountability and Policies & 32\% \\
      Ethical Frameworks and Moral Reasoning & 24\% \\
      Human-AI Interaction and Agency & 24\% \\
      Algorithmic Bias and Fairness & 18\% \\
      AI Safety and Robustness & 16\% \\
      Generative AI specific concerns & 11\% \\
      Social Impact and Justice & 8\% \\ 
        \hline
        \end{tabular} \vspace{-0.35cm}
\end{table}

\textbf{Privacy, Legal aspect and data governance}, covered by 42\% of courses, dealt with responsible data use, anonymization, and compliance with privacy frameworks. In addition to general privacy principles, several courses explored specific concerns such as privacy risks in generative AI and large language models. Other topics included algorithmic approaches for protecting personal traits from being reverse-engineered and discussions around the ethical use of clinical data in healthcare machine learning systems. Many courses also addressed broader policy and legal frameworks for regulating AI privacy, Regulatory challenges in AI, International efforts for AI governance, Ethical considerations in AI policy-making Ethical Decision-Making in AI: Ethical decision-making frameworks, Case studies on ethical dilemmas in AI, Ethical impact assessments Future Directions and Emerging Ethical Challenges: Emerging ethical challenges in AI: Ethical considerations in advanced AI technologies, Ethical leadership in AI development and deployment.

More specialized but still significant themes include \textbf{accountability and policy} (32\%), which explores questions of responsibility in AI system failures and oversight structures. Courses addressing this theme cover topics such as responsible AI deployment and the integration of ethical principles into policy frameworks, often illustrated through case studies in healthcare and education. 

Meanwhile, \textbf{ethical frameworks and moral reasoning} are covered in 24\% of the courses, introducing students to philosophical approaches and equipping them to reason through complex dilemmas posed by intelligent systems. These frameworks are applied to AI-related moral dilemmas, helping students develop skills in ethical reasoning and decision-making. Topics include ethical considerations in digital technologies and real-world case studies that challenge students to connect theoretical concepts with contemporary AI issues.
Few handouts specifically mention ethical impact on society, impact on human psychology, impact on the legal system, impact on the environment and the planet, and impact on trust.

Courses addressing the theme of \textbf{human-AI interaction and agency} (24\%) focus on how AI systems can be designed to support meaningful, ethical, and effective collaboration between humans and machines. Topics include human-centered design principles, user interface strategies for AI-driven systems, and methods for enabling human-AI teaming and interoperability

The topic \textbf{algorithmic bias and fairness} in intelligent systems is covered in 18\% of courses. These topics introduce students to methods for detecting bias in data and model outputs, highlighting how social inequalities can be encoded and amplified by machine learning systems. Several courses go further by covering post-processing strategies, often grounded in real-world case studies. The ethical implications of biased systems are also emphasized, particularly in high-stakes applications such as healthcare, medical algorithms, robots, Autonomous Vehicles, Warfare, and weaponization.

The topic of \textbf{AI safety and robustness} (16\%), which focuses on reliability, adversarial robustness, and secure system behavior in real-world settings. Topics covered include safe reinforcement learning, as well as emerging risks posed by generative AI and large language models.

\textbf{Generative AI-specific concerns}—including misinformation, value misalignment, and prompt injection—appear in 10.67\% of courses. Courses also engage with the societal implications of generative AI, including misinformation, content moderation, privacy concerns and regulatory challenges. Finally. Some courses discuss sustainable system design and advocate for ethical awareness of the ecological and human toll of AI development.

Courses also address topics around \textbf{social impact and justice} (8\%), focusing on how AI technologies influence domains such as healthcare, education, social justice, law enforcement, and employment. These courses encourage students to consider the equity implications of AI deployment, especially on marginalized communities. Specific topics include the role of AI in predictive policing and public safety, as well as understanding systemic inequities embedded in data-driven decision systems. Some courses integrate frameworks of justice, rights, and welfare into machine learning curricula and emphasize participatory or interdisciplinary design involving experts from public health. \vspace{-0.35cm}

\subsubsection{Resources used to teach AI Ethics}
We examined the syllabi for references to AI ethics-specific resources or textbooks. The majority of the courses (over 90\%) did not include any dedicated materials for teaching AI ethics in the syllabus. Instead, they primarily included standard AI and machine learning textbooks—such as Artificial Intelligence: A Modern Approach~\cite{10.5555/1671238}, Pattern Recognition and Machine Learning~\cite{10.5555/1162264}, and other widely used texts aligned with the technical focus of the core course in which AI ethics content was embedded. Only a small number of syllabi explicitly mentioned ethics-specific resources, such as AI for Social Impact\footnote{\url{https://ai4sibook.org/}}, Towards a Code of Ethics for Artificial Intelligence by Paula~\cite{paula_boddington_towards_2017}, AI Ethics by Mark~\cite{mark_coelcerkberg_ai_2020}, and Responsible Artificial Intelligence: How to Develop and Use AI in a Responsible Way by Dignum~\cite{dignum_responsible_2019} and The Oxford Handbook of Ethics of AI~\cite{dubber_oxford_2020}. \vspace{-0.25cm}

\subsection{Learning Objectives (LO) in AI Ethics Education}
This section presents the learning objectives (LO), assessment methods found from our analysis.   \vspace{-0.25cm}

\subsubsection{Learning Objectives}
The terms “learning objectives” and “learning outcomes” have subtle differences in meaning but are often used interchangeably~\cite{Harden01012002}; we saw both in our data. Similar to a 2020 syllabus analysis published in SIGCSE on tech ethics~\cite{fiesler_what_2020}, we use the term \textit{learning outcome} to mean an explicit statement of what students should know or be able to do by the end of the course related to AI ethics, regardless of how such a statement might be labeled in any given syllabus. Our qualitative findings revealed eight commonly stressed upon types of learning outcomes in terms of what \textit{students should be able to} do by the end of the course. The LO found are: 
\begin{enumerate}
    \item Identify and evaluate privacy risks in AI systems, and propose strategies to mitigate them within ethical and legal frameworks
    \item ethical considerations: fairness in AI (bias, transparency, implications of AI decisions on society).
    \item Interpret and communicate the decision-making processes of AI systems using explainability techniques
    \item Discuss or Critique key ethical dilemmas encountered during the design, training, and deployment choices of AI systems
    \item Investigate and compare ethical frameworks, regulatory strategies, and policy approaches relevant to AI governance and AI healthcare
    \item Apply ethical reasoning models to assess and justify decisions made in the development and deployment of AI systems
    \item Policy for AI ethics
    \item Global governance on AI
\end{enumerate} \vspace{-0.5cm}

\subsubsection{Assessment Methods}
None of the courses included documented assessments explicitly dedicated to AI ethics topics. In cases where assessments were mentioned, they were limited to broad categories—such as quizzes, assignments, projects, mid-semester, or end-semester exams—without specifying the content or indicating whether any of these components addressed AI ethics. While it is possible that ethics-related questions may be embedded within broader evaluations (say in written exams), such details were generally absent from the syllabi reviewed. These findings suggest that the assessment of AI ethics remains limited, lacking both clarity and consistent integration within computing curricula. \vspace{-0.5cm}

\subsection{Miscellaneous Findings} 
Notably, out of the top 100 engineering institutions identified through the National Institutional Ranking Framework (NIRF), fewer than half had publicly accessible academic curricula, course handouts, or detailed syllabus information available on their official university websites. 
Interestingly, one private institution offering a dedicated AI/ML engineering degree did not include any course that addressed AI ethics in any form. While this observation is not necessarily indicative of the absence of such content in practice, it highlights the limited visibility of ethics-related coursework in program documentation. This lack of transparency raises important questions about the extent to which ethical considerations are being systematically integrated into the design of emerging AI curricula.\vspace{-0.35cm}

%% file: discussion.tex
\section{Discussion and Conclusion}
\label{sec:Discussions}

This study presents the first large-scale analysis of how AI ethics is represented in computing education across top Indian engineering institutions. Despite the increasing global emphasis on ethical AI development, our findings reveal that AI ethics remains minimally integrated and inconsistently represented in Indian computer science and allied curricula.

Only 2.21\% of the 3,395 courses analyzed included any substantive AI ethics content, and the majority of those addressed it superficially—often through a single lecture session. Courses offering in-depth or dedicated instruction were rare, and even fewer articulated clear learning outcomes or assessment methods tied to ethical topics.

Our findings align closely with the ACM/IEEE/AAAI Computer Science Curricula 2023 (CS2023)~\cite{10.1145/3664191}, which identifies AI, Society, and Professionalism as a key knowledge area containing several CS Core competencies. Many of the learning outcomes (LOs) identified in our analysis— such as privacy, accountability, and fairness, policies, ethical frameworks, and moral reasoning —correspond well with the ethical and societal objectives outlined in CS2023. However, despite this encouraging overlap, important gaps remain. Notably, there is limited curricular attention to the ethical challenges posed by deep generative models, such as large language models and deepfakes, which are explicitly highlighted in CS2023 as essential topics. Only 11\% of courses (N=75) addressed these issues. Similarly, sustainability, justice, and labor ethics—including the environmental costs of model training were covered in just 8\% of courses. These omissions point to a disconnect between current teaching practices and the broader vision of socially responsible computing promoted in CS2023, underscoring the need for more comprehensive and future-facing AI ethics education.

A significant gap was identified in the availability and use of instructional resources: most syllabi relied on general AI or machine learning textbooks, with minimal or no reference to ethics-specific materials. Additionally, assessment practices related to AI ethics were either vaguely described or entirely missing, raising concerns about how students' understanding of ethical issues is being evaluated. These trends point to a broader lack of structured pedagogical design for AI ethics education and reflect inconsistencies in how instructors conceptualize and incorporate ethical content into their courses. The absence of widely adopted teaching materials likely stems from uncertainty about what constitutes core ethical knowledge in AI, along with a shortage of accessible, well-designed resources. There is, therefore, a pressing need to develop comprehensive textbooks, teaching guides, and assessment frameworks that can support educators in meaningfully integrating AI ethics into computing curricula. This also presents an opportunity to contextualize AI ethics education within issues of particular relevance to India, such as algorithmic bias linked to caste and religion, the ethics of large-scale biometric systems like Aadhaar, and the societal implications of AI in agriculture and public distribution systems. A useful comparison can be drawn from accessibility education in computer science, which, until recently, was similarly underrepresented. The development of dedicated resources and textbooks\footnote{\url{https://bookish.press/tac}
, \url{https://accessibilityeducation.github.io/}}—complete
 with practical examples and assessment tools—has significantly advanced its curricular adoption. A parallel effort is now essential in AI ethics to enable consistent, rigorous, and scalable integration across academic institutions.

The limited availability of publicly accessible curricula posed a significant barrier to conducting a truly comprehensive review across all top institutions. This lack of transparency in curriculum sharing reflects broader structural challenges in accessing organized educational data within the Indian higher education system—particularly when assessing the integration of emerging topics like AI ethics. In contrast, universities in the Global North, especially in countries like the U.S. and across Europe, commonly publish detailed course syllabi and learning outcomes online as part of a broader commitment to openness, accountability, and academic exchange. The relative absence of such practices in India highlights a critical need for greater institutional support for open curricular access. Enhancing transparency in curriculum dissemination would not only facilitate educational research and policy evaluation but also foster cross-institutional learning and capacity-building for AI ethics instruction.

Overall, our findings reveal a fundamental concern: AI ethics remains a marginal and unevenly addressed component of CS education. Despite increasing discourse on the ethical responsibilities of AI professionals, this awareness has yet to be reflected in the structure, content, and pedagogy of most CS programs. The absence of shared curricular frameworks, reliable instructional resources, and clearly defined assessment strategies contributes to a fragmented landscape in which ethics is often sidelined or superficially integrated. To move beyond this piecemeal approach, there is an urgent need for coordinated efforts that embed AI ethics as a core element of computing education. This entails establishing consensus-driven learning goals, developing high-quality open educational materials—including modular textbooks, practical case studies grounded in the Indian context, and scaffolded activities—and designing assessments that go beyond rote evaluation to cultivate ethical judgment and reflective thinking. Additionally, empowering faculty through professional development opportunities and cross-disciplinary collaborations will be essential, especially for instructors new to ethics instruction. %Engaging the computing education community in forums like iSIGCSE\footnote{\url{https://isigcse.acm.org/index.html}}, EdTech society\footnote{\url{https://etsociety.org/}} to co-develop shared norms and expectations can help build a collective foundation for AI ethics education. 
Without such systemic efforts, ethical understanding will remain an afterthought, rather than a foundational skill for the next generation of AI practitioners. \vspace{-0.35cm}

\section{Threats to Validity}
\label{sec:threats} \vspace{-0.25cm}
While this study provides a systematic analysis of AI ethics coverage in computing curricula, several limitations must be acknowledged. First, although efforts were made to manually collect the most recent and publicly available syllabi from top-ranked institutions, the process was subject to human error, especially when navigating decentralized or outdated university websites. Some relevant syllabi may have been missed due to inconsistencies in how courses are titled, categorized, or published online. Second, our sample focused exclusively on the top 100 engineering institutions as per the National Institutional Ranking Framework (NIRF). While this offers a robust view of leading programs, it does not capture the full spectrum of institutions—particularly smaller or unranked colleges—that may also be offering meaningful instruction in AI ethics. Future work could expand the sample and use direct faculty surveys or institutional outreach to validate and enrich the findings.